\documentclass[12pt]{article}
\usepackage{axodraw}
\usepackage{pstricks}
\usepackage{epsfig}
\newcommand{\singlespace}{\renewcommand{\baselinestretch}{1}\large\normalsize}

\begin{document}

\begin{titlepage}
\pagestyle{empty}
\setcounter{page}{1}
\vspace{1.0in}

\begin{center}
\singlespace
\begin{Large}
{\bf Anomalous decay of pion and eta at finite density}\\
\end{Large} 
\end{center}
\vskip 2cm
\begin{center}
{\large {P. Costa}}\footnote{pcosta@teor.fis.uc.pt}, 
{\large {M. C. Ruivo}}\footnote{maria@teor.fis.uc.pt}

{\em  Departamento de F\'{\i}sica
da Universidade de Coimbra

P - 3004 - 516 Coimbra, Portugal}

\vspace{0.3cm}

{\large {Yu. L. Kalinovsky}}\footnote{kalinov@nusun.jinr.ru}

{\em  Laboratory of Information Technologies,

Joint Institute for Nuclear Research, Dubna, Russia}
\end{center}
\vspace{2cm}

\begin{abstract}
We study the anomalous decays $\pi^0,\, \eta \rightarrow\gamma\gamma$ in
the framework of the three--flavor Nambu--Jona-Lasinio [NJL] model, in the vacuum 
and in quark matter in $\beta$ --equilibrium. It is found that the behavior of the 
relevant observables essentially reflects a manifestation of the partial restoration 
of chiral symmetry, in non strange and strange sectors. The probability of such 
decays   decreases   with density, showing  that anomalous mesonic interactions are significantly affected  by the medium. 
\bigskip


\end{abstract}
\end{titlepage}
\newpage

%%%%%%%%%%%%%%%%%%%%%%%%%%%%%%%%%%%%%%%%%%%%%%%%%%%%%%%%%%%%%%%%%%%%%%%%%%%%%%%%%%%%%%%%%%%%%%%%%%%%%%%%%%%%%%%%%%%%%%%%

\section{Introduction}
\hspace*{\parindent} The structure of pseudoscalar mesons, its mass spectra and decays have  attracted a lot of interest along the years, an important motivation for this interest being certainly related to the fact that the origin of these mesons is related to the spontaneous or explicit breakdown of symmetries of QCD.
Furthermore,  since at high densities and temperatures new phases with restored symmetries are expected to occur, the study of pseudoscalar meson observables under those conditions is specially relevant since they could provide signs for the phase transitions and the associated restoration of symmetries. As a matter of fact, major theoretical and experimental efforts  have been dedicated to heavy--ion physics looking for signatures of the quark gluon plasma, a state of matter with deconfinement of quarks and restoration of symmetries  \cite{kanaya,rhic,cern}.

In the limit of vanishing quark masses, the QCD Lagrangian has 8 Goldstone bosons,
associated with the dynamical breaking of chiral symmetry. The non inexistence of a ninth Goldstone boson in QCD  is explained by assuming that    the QCD Lagragian
has a $U_A (1)$ anomaly.  The origin of the physical masses of the different pseudoscalar mesons is due to the explicit breaking of symmetries, but it presents differences that will be analyzed next. While for  pions and kaons it is enough to break explicitly the chiral symmetry  by giving current masses to the quarks, the  breaking of the $U_A (1)$ symmetry by instantons has the  effect of giving a mass
to $\eta'$ of about $1$ GeV. So the mass of $\eta'$ has a different origin
than the masses of the other pseudoscalar mesons and it cannot be regarded
as the remnant of a Goldstone boson. Concerning the other two neutral mesons, $\eta$ and $\pi^0$, they are degenerated when the $U_A(1)$ anomaly and the current quark masses are turned off, but, when these effects are taken into account it turns out that  a percentage of $\eta$
mass is due to the anomaly, and, therefore,  this meson should regarded as  less "Goldstone--like" than the pion. 
Investigating this problem in vacuum and in medium is an important  task.  
Besides mass spectrum and meson--quark coupling constants, the  observables 
associated with the two photon decay of these mesons might  provide 
useful insight into this problem, and  calculations of  such observables 
in vacuum and at finite temperature may be found in the literature 
\cite{cleo,tdavid,tklev,tpisarski,pisarski,dorokhov,oka,hashimoto}.

Understanding the processes $\pi^0 (\eta)\,\rightarrow \gamma\gamma$ is specially relevant
having in mind that the great percentage of photons in the background of heavy-ion collisions is due to the decay of $\pi^0\mbox{and}\, \eta$ \cite{star}. As a matter of fact the production of such mesons is indicated by the occurrence of two photon pairs with an invariant mass equal to these meson masses. Possible medium modifications of anomalous mesonic interactions  is a topic that has attracted lost of interest.  As pointed by Pisarski \cite{tpisarski} and Pisarski and Tytgat \cite{pisarski} while for fermions the axial anomaly is not affected by the medium, the opposite is expected for anomalous mesonic interactions. In this concern, the study of $\pi^0\,\rightarrow \gamma\gamma$ is particularly interesting due to its simplicity and its association with restoration of chiral symmetry. Although the lifetime of the neutral pion is much longer than hadronic time scales and it is not expected to be observed inside the fireball, the physics is the same as that of other anomalous decays ($\omega\rightarrow \pi\pi\pi\,, \omega \rightarrow \rho\pi$) that are relevant for experiments in the hot/dense region.

A great deal of knowledge on chiral symmetry breaking and restoration, as well as on meson properties, comes from model calculations \cite{roberts}. In particular, the Nambu--Jona-Lasinio [NJL] \cite{NJL,Volkov}  type models have been extensively used
over the past years to describe low energy features of hadrons and also to
investigate restoration of chiral symmetry with temperature or density 
\cite{njlt,Hatsuda94,RuivoSousa,RSP,Ruivo,RKH,ruivoI,costaI,costaB,costabig}. 

This work comes in the sequel of previous studies on the behavior of neutral mesons in hot and dense matter \cite{costaI,costaB,costabig}. The study of phase transitions in quark matter simulating neutron matter in $\beta$ equilibrium, at zero and finite temperature, as well as the discussion of behavior of pseudoscalar mesons (in particular neutral mesons) in such media, in connection with the restoration of symmetries, has been done in \cite{costaI,costaB,costabig},  by analyzing only the mass spectrum. In particular, it was shown that the behavior of the masses of $\pi^0$ and $\eta$ in this concern reflects mainly the restoration of chiral $SU(2)$ symmetry  and  manifests  strongly in the pion. This not so evident for the $\eta$, due to its strangeness content and the dependence of the anomaly.  
The aim of this paper is to investigate the anomalous decays of $\pi^0$ and $\eta$ 
in the vacuum and in quark matter in $\beta$ --equilibrium and discuss our 
results in connection with the breaking and restoration of symmetries. We also compare our results with those obtained by other authors who studied the effect of temperature on these anomalous mesonic interactions \cite{tdavid,tklev,tpisarski,hashimoto}.

%%%%%%%%%%%%%%%%%%%%%%%%%%%%%%%%%%%%%%%%%%%%%%%%%%%%%%%%%%%%%%%%%%%%%%%%%%%%%%%%%%%%%%%%%%%%%%%%%%%%%%%%%%%%%%%%%%%%%%%%

\section{Formalism}
    
\hspace*{\parindent} We consider the three--flavor  NJL type model containing  
scalar--pseudoscalar interactions and a determinantal term, the 't Hooft
interaction, with the following Lagrangian: 
\begin{equation}
\begin{array}{rcl}
{\mathcal L\,} & = & \bar q\,(\,i\, {\gamma}^{\mu}\,\partial_\mu\,-\,\hat m)\, q+%
\frac{1}{2}\,g_S\,\,\sum_{a=0}^8\, [\,{(\,\bar q\,\lambda^a\, q\,)}%
^2\,\,+\,\,{(\,\bar q \,i\,\gamma_5\,\lambda^a\, q\,)}^2\,] \\[4pt] 
& + & g_D\,\, \{\mbox{det}\,[\bar q\,(\,1\,+\,\gamma_5\,)\,q\,] + \mbox{det}%
\,[\bar q\,(\,1\,-\,\gamma_5\,)\,q \,]\, \}. \label{1} \\ 
&  & 
\end{array}
\label{lag}
\end{equation}
Here $q = (u,d,s)$ is the quark field with three flavors, $N_f=3$, and
three colors, $N_c=3$. $\hat{m}=\mbox{diag}(m_u,m_d,m_s)$ is the current 
quark masses matrix and $\lambda^a$ are the Gell--Mann matrices, 
a = $0,1,\ldots , 8$, ${ \lambda^0=\sqrt{\frac{2}{3}} \, {\bf I}}$.

The last term in (\ref{lag}) is the lowest six--quark dimensional 
operator and it has the $SU_L(3)\otimes SU_R(3)$ invariance but breaks the
$U_A(1)$ symmetry. This term is a reflection of the axial anomaly in QCD and
can be put in a form suitable to use the
bosonization procedure (\cite{Ruivo,costaB} and references therein):
\begin{equation}
{\mathcal L_D}\,=\, \frac{1}{6} g_D\,\, D_{abc} \,(\bar q\, {\lambda}%
^c\,q\,)\,[\,(\,\bar q\,\lambda^a\, q\,)(\bar q\,\lambda^b\, q\,) -
3\,(\,\bar q \,i\,\gamma_5\,\lambda^a\, q\,)\,(\,\bar q
\,i\,\gamma_5\,\lambda^b\, q\,)\,]
\end{equation}
with constants $D_{abc}=d_{abc}$ if $a,b,c  \in \{1,2,..8\}\,$, where $d_{abc}$ 
are the $SU(3)$  structure constants,  and $D_{000}=\sqrt{\frac{2}{3}}$,
$D_{0ab}=-\sqrt{\frac{1}{6}}\delta_{ab}$.

The usual procedure to obtain a four quark effective interaction from the
six quark interaction is to contract one bilinear $(\bar q\,\lambda_a\,q)$ 
making a shift $(\bar{q}\lambda^a q) \rightarrow (\bar{q}\lambda^a q) +
<\bar{q}\lambda^a q>$ with the vacuum expectation value $<\bar{q}\lambda^a q>$.
Then, an effective Lagrangian  can be written  as:
\begin{eqnarray}
{\mathcal L}_{eff} &=& \bar q\,(\,i\, {\gamma}^{\mu}\,\partial_\mu\,-\,\hat m)\, q \,\,
\nonumber \\
&+& S_{ab}[\,(\,\bar q\,\lambda^a\, q\,)(\bar q\,\lambda^b\, q\,)]
+\,P_{ab}[(\,\bar q \,i\,\gamma_5\,\lambda^a\, q\,)\,(\,\bar q
\,i\,\gamma_5\,\lambda^b\, q\,)\,],
\label{lagr_eff}
\end{eqnarray}
where
\begin{eqnarray}
S_{ab}\,=\,g_S\,\delta_{ab}\,+\,g_D D_{abc} \,< \bar q\, {\lambda}^c\,q\,>,
\nonumber \\
P_{ab}\,=\,g_S\,\delta_{ab}\,-\,g_D D_{abc} \,< \bar q\, {\lambda}^c\,q\,>
\label{pab}.
\end{eqnarray}
Integration over quark fields in the functional integral with (\ref{lagr_eff})
gives the meson effective action  
\begin{eqnarray}  \label{action}
W_{eff}=&-i& \mbox{Tr} \ {\rm ln}(\,i\,\partial_\mu \gamma_{\mu}-\hat
m+\sigma_a\,\lambda^a +i\,\gamma_5\, {\phi}_a\,\,{\lambda}^a\,)  \nonumber \\
&-&\frac{1}{2}(\,\sigma_a\,{S_{ab}}^{-1}\,\sigma_b\,+{\phi}_a\, {\ P_{ab}}%
^{-1}\,\phi_b\,).
\end{eqnarray}
The fields $\sigma^a$ and $\phi^a$  are the  scalar and pseudoscalar 
meson nonets.

The first variation of the action (\ref{action}) leads to the set of 
gap equations for constituent quark masses $M_i$:
\begin{equation}
M_i\,=\,m_i\,-2\,g_S\,<\bar q_i\, q_i>\,-\,2\,g_D\,<\bar q_j\, q_j><\bar
q_k\, q_k> \, 
\end{equation}
with $i\,,j\,,k = u,d,s$  cyclic and $ < \bar{q}_i\, q_i> = - i \mbox{Tr} S_i(p) $ 
are the quark condensates. Here the symbol $\mbox{Tr}$ 
means trace in color and Dirac spaces and integration over momentum $p$ with 
a cut--off parameter $\Lambda$ to regularize the divergent integrals. 
The pseudoscalar meson masses  $M_H$ ($H =\pi^0,\,\eta$)  are obtained  from the condition 
$(1 - P_{ij}\, \Pi^{ij}(P_0=M_H,\textbf{P}=0) ) = 0$, where 
$\Pi^{ij}(P_0=M_H,\textbf{P}=0)$ is the polarization operator at the rest frame  
\begin{equation}
\Pi^{ij}(P_0)=4 \left[ (I^i_1 + I^j_1) - 
\left( P_0^{2} -(M_i -M_j)^2 \right) I_{2}^{ij}(P_0) \right]\,,
\end{equation}
and  integrals are given by
\begin{equation}
I^i_{1}=\frac{N_{c}}{4\pi^{2}}\int^{{\Lambda }}_{0}%
\frac{{\tt p}^{2}}{E_{i}}d{\tt p},  \label{I1} 
\end{equation}
\begin{equation}
I_{2}^{ij}(P_0) =\frac{N_{c}}{4\pi ^{2}} \int^{\Lambda }_{0}
\frac{{\tt p}^{2} d{\tt p}}{E_i E_j} \,\, 
\frac{E_i+E_j}{P_0^{2}-(E_i+E_j)^2} 
+i  \frac{1}{2\pi} \,\,
\frac{p^*}{(E_i^*+E_j^*)} \, ,
\label{I2}
\end{equation}
where $E_{i,j}= \sqrt{{\tt p}^2 +M_{i,j}^2}$ and  
$E_{i,j}^*= \sqrt{({\tt p}^*)^2 +M_{i,j}^2}$ are the quark energies. The momentum  
$p^* $ is defined by  $p^* = \sqrt{(P_0^{2}-(M_i-M_j)^2)(P_0^{2}-(M_i+M_j)^2)}/2P_0$.

The quark--meson coupling constant is evaluated as  
\begin{equation}
g_{H\bar{q}q}^{-2}=%
-\frac{1}{2 M_H}\frac{\partial}{\partial P_0}[\Pi_{ij}(P_0)]_{|P_0=M_H},
\label{coupl}
\end{equation}
where the bound state contains quark flavors $i,j$.

Having the on--shell quark--meson coupling constant we can calculate the 
meson decay constant $f_H$ according to the definition 
\begin{eqnarray}
f_H = N_c g_{H\bar{q}q} \frac{P_\mu}{P^2} 
\int \frac{d^4 p}{(2\pi)^4}
\mbox{tr} \left[ (i \gamma_5) S_i(p) (\gamma_\mu \gamma_5) S_j(p+P) \right].
\end{eqnarray}

Our model parameters, the bare quark masses $m_d=m_u, m_s$, 
the coupling constants and the cutoff in three--momentum space, $\Lambda$, 
are  fitted to the experimental values of masses for
pseudoscalar mesons ($M_{\pi^0} = 135.0$ MeV, $M_{K} = 497.7$) and 
$f_\pi = 92.4$ MeV. 

Here we use the following  parameterization \cite{RKH,costaB,costabig}: 
$\Lambda=602.3$ MeV, 
$g_S \Lambda^2=3.67$, 
$g_D\Lambda^5=-12.39$, 
$m_u=m_d=5.5$ MeV and 
$m_s=140.7$ MeV.

We also have $M_\eta = 514.8$ MeV,
$\theta (M_\eta^2) = -5.8^{\circ}$,
$g_{\eta\bar{u}u} = 2.29$, 
$g_{\eta\bar{s}s} = -3.71$.  

Note, that $\theta (M_\eta^2)$ is the mixing angle which represents
the mixing  of $\lambda^8$ and $\lambda^0$ components in the 
$\eta$ --meson state (for details see \cite{costaB,costabig}).

For the quark condensates we have: 
$<\bar{u}u>\,\,\,=\,\,\,<\bar{d}d>\,\,\,=\,\,\,-(241.9 \mbox{MeV})^3$,
$<\bar{s}s>=-(257.7 \mbox{MeV})^3$, and 
$M_u = M_d = 367.7$ MeV, $M_s = 549.5$ MeV, for the constituent       
quark masses. 
In this section we have described the NJL model with the 't Hooft determinant.
The model describes well the vacuum properties related with chiral symmetry and 
its spontaneous breaking  including their flavor dependence.

%%%%%%%%%%%%%%%%%%%%%%%%%%%%%%%%%%%%%%%%%%%%%%%%%%%%%%%%%%%%%%%%%%%%%%%%%%%%%%%%%%%%%%%%%%%%%%%%%%%%%%%%%%%%%%%%%%%%%%%%

\section{The decay $H \longrightarrow \gamma \gamma$}

\hspace*{\parindent}For the description of the decays $H \longrightarrow \gamma \gamma$ we consider 
the triangle diagrams for the electromagnetic meson decays. They are shown 
in Fig. 1. The corresponding invariant amplitudes are given by 
%%%%%%%%%%%%%%%%%%%%%%%%%%%%%%%%%%%%
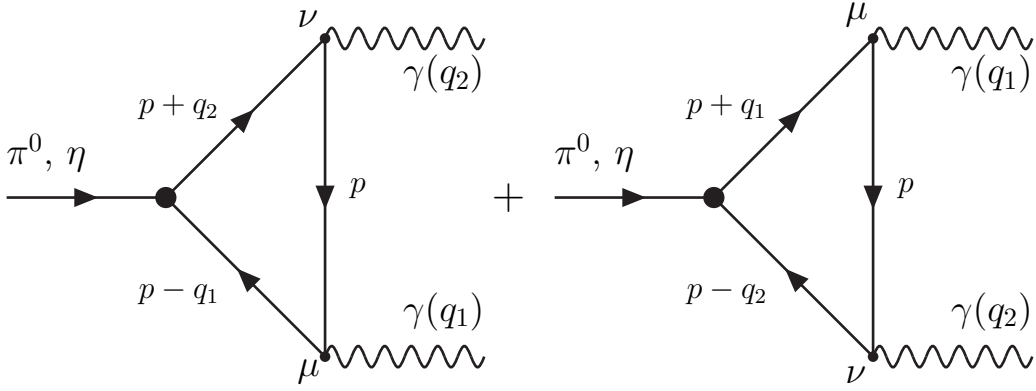
\begin{figure}[t]
\[
\begin{picture}(125,86)(80,80)

\hspace*{2cm}
\SetScale{2.0}
{
\ArrowLine(0,30)(30,30)
\ArrowLine(30,30)(60,60)
\ArrowLine(60,60)(60,0)
\ArrowLine(60,0)(30,30)
\Vertex(30,30){2}
\Vertex(60,0){1}
\Vertex(60,60){1}
\Photon(60,60)(90,60){2}{6}
\Photon(60,0)(90,0){2}{6}
\Text(130,60)[lb]{$p$}
\Text(50,90)[lb]{$p+q_2$}
\Text(50,20)[lb]{$p-q_1$}
\Text(150,100)[lb]{\large $\gamma (q_2)$}
\Text(150,10)[lb]{\large $\gamma (q_1)$}
\Text(0,70)[lb]{\large $\pi^0$, $\eta$}
\Text(110,125)[lb]{\large $\nu$}
\Text(110,-10)[lb]{\large $\mu$}
}
\end{picture}
\begin{picture}(125,86)(80,80)
\hspace*{1.25cm}
\SetScale{2.0}
\Text(80,55)[lb]{\Large $+$}
\end{picture}
\begin{picture}(125,86)(80,80)
\hspace*{0.5cm}
\SetScale{2.0}
{
\ArrowLine(0,30)(30,30)
\ArrowLine(30,30)(60,60)
\ArrowLine(60,60)(60,0)
\ArrowLine(60,0)(30,30)
\Vertex(30,30){2}
\Vertex(60,0){1}
\Vertex(60,60){1}
\Photon(60,60)(90,60){2}{6}
\Photon(60,0)(90,0){2}{6}
\Text(130,60)[lb]{$p$}
\Text(50,90)[lb]{$p+q_1$}
\Text(50,20)[lb]{$p-q_2$}
\Text(150,100)[lb]{\large $\gamma (q_1)$}
\Text(150,10)[lb]{\large $\gamma (q_2)$}
\Text(0,70)[lb]{\large $\pi^0$, $\eta$}
\Text(110,125)[lb]{\large $\mu$}
\Text(110,-10)[lb]{\large $\nu$}
}
\end{picture}
\]
\vspace{2.5cm}\caption{The quark triangle diagram for the 
$H\rightarrow\gamma\gamma$ (direct and exchange process).}
\label{fig:direct1}
\end{figure}
%%%%%%%%%%%%%%%%%%%%%%%%%%%%%%%%%%%%
\begin{eqnarray}
{\tilde{\mathcal T}}_H(P,q_1,q_2) &=& 
 i  \int \frac{d^4 p}{(2\pi)^4}
\textrm{Tr} \left\{ \Gamma_H S(p - q_1) 
\hat{\epsilon}_1 
S(p)  \hat{\epsilon}_2 
S(p + q_2) \right\} \nonumber \\ &&
+ \textrm{exchange}. \label{trian}
\end{eqnarray}
Here the trace
$\textrm{Tr}=\textrm{tr}_c\textrm{tr}_f\textrm{tr}_\gamma$, 
must be performed over color, flavor  and spinor indices.
The meson vertex function $\Gamma_H $  has the $i \gamma_5$ form 
in the Dirac space, contains the corresponding coupling constant $g_{H\bar{q}q}$ 
(see (\ref{coupl})) and presents itself the $3 \times 3$ matrix form in the flavor space. 
 $S(p)$ is the quark propagator $S(p)=\mbox{diag}(S_u, S_d, S_s)$, 
$\hat{\epsilon}_{1,2}$ is the photon polarization vector with momentum $q_{1,2}$.
The trace over flavors leads to different factors for different mesons $H$: $Q_{H\bar{q}q}$. 
This factor depends on the electric charges and flavor of quarks into the meson $H$: 
$Q_\pi=1/3$, $Q_{\eta_u}=5/9$ and $Q_{\eta_s}=-\sqrt{2}/9$. 

For this evaluation, we move to the meson rest frame and use the kinematics $P=q_1+q_2$ and $P=(M_{H},\textbf{0})$.
Taking the trace 
in (\ref{trian}) we can obtain
\begin{eqnarray}
    {\tilde{\mathcal T}}_H(P,q_1,q_2) = \epsilon_{\mu\nu\alpha\beta}\epsilon^\mu_1\epsilon^\nu_2
    q_1^\alpha q_2^\beta  \, {\mathcal T}_H (P^2,q_1^2,q_2^2) 
\end{eqnarray}
where 
\begin{eqnarray}
    {\mathcal T}_{\pi^0} (P^2=M_{\pi^0}^2,q_1^2,q_2^2) =
    32\alpha\pi g_{\pi^0\bar{u}u} 
    I^u_{\pi^0}  
\end{eqnarray}
and
\begin{eqnarray}
	& &{\mathcal T}_{\eta} (P^2=M_{\eta}^2,q_1^2,q_2^2) = \nonumber \\
		& &\frac{32\alpha\pi}{3\sqrt{3}}\bigl[  
		\mbox{cos} \theta  (5 g_{\eta\bar{u}u}I_\eta^u - 2 g_{\eta\bar{s}s}I_\eta^s )
		- \mbox{sin} \theta  \sqrt{2} (5 g_{\eta\bar{u}u}I_\eta^u + g_{\eta\bar{s}s}I_\eta^s )
	\bigr]. 
\end{eqnarray}
where $\alpha$ is the fine structure constant. 
The integrals $I_{H}^i\equiv I_{H}^i(P) $ are  given by
\begin{equation} \label{inth}
I_{H}^i(P)=iM_{i}\int\frac{d^{4}p}{(2\pi)^{4}}\frac{1}{(p^{2}-M_{i}^{2})%
[(p-q_1)^{2}-M_{i}^{2}][(p+q_2)^{2}-M_{i}^{2}]}.
\end{equation}
In order to introduce finite density effects, we apply the Matsubara technique 
\cite{tklev,kapusta}, and the integrals relevant for our calculation ((\ref{I1})
and (\ref{I2})) are then modified in a standard way \cite{costaI,costaB,costabig}. 
In particular $I_H^i(P)$ (\ref{inth}) takes the form: 
\begin{equation}
I_{H}^i(P_0,\textbf{P}=0)=-\frac{M_{i}}{4\pi^2}
\int_{\lambda_i}^{\infty}dp\frac{p}{E_{i}^2}\frac{1}{4E_{i}^2-P_0^2}\textrm{ln}(\frac{E_{i}+p}{M_i})  \label{i3}
\end{equation}
where $\lambda_i$ is the Fermi momentum.

Finally, the decay width is obtained from
\begin{equation}
\Gamma_{H\rightarrow\gamma\gamma}=\frac{M_{H}^{3}}{64\pi}|\mathcal{T}_{H\rightarrow\gamma\gamma}|^{2}%
\end{equation}
and the decay coupling constant is
\begin{equation}
g_{H\rightarrow\gamma\gamma}=\frac{\mathcal{T}_{H\rightarrow\gamma\gamma}}{e^2}.%
\end{equation}

%%%%%%%%%%%%%%%%%%%%%%%%%%%%%%%%%%%%%%%%%%%%%%%%%%%%%%%%%%%%%%%%%%%%%%%%%%%%%%%%%%%%%%%%%%%%%%%%%%%%%%%%%%%%%%%%%%%%%%%%

\section{Discussion and conclusions} 

\hspace*{\parindent} We present in Table I our results for $\mathcal{T}_{H\rightarrow\gamma\gamma}$, 
$\Gamma_{H\rightarrow \gamma \gamma}$ and $g_{H\gamma \gamma}\, $ 
($H=(\pi^0\,,\eta)$) in the vacuum, in comparison with experimental results 
\cite{cleo,dorokhov,pdb}, and we can see that there is a good agreement. 
\\
\begin{table}[h]
\caption{Comparison of the experimental values with numerical results 
obtained in the NJL model.}
\begin{center}
    \begin{tabular}
    [c]{cc|c|c|}\cline{3-4}%
    &  & NJL & Exp.\\\hline\hline
    \multicolumn{1}{|c}{} & \multicolumn{1}{|c|}{$|\mathcal{T}_{\pi^{0}%
    \rightarrow\gamma\gamma}|\,[$eV$]^{-1}$} & $2.5\times10^{-11}$ & $(2.5\pm
    0.1)\times10^{-11}$\\\cline{2-4}%
    \multicolumn{1}{|c}{$\pi^{0}$} & \multicolumn{1}{|c|}{$\Gamma_{\pi^{0}%
    \rightarrow\gamma\gamma}\,[$eV$]$} & $7.65$ & $7.78(56)$\\\cline{2-4}%
    \multicolumn{1}{|c}{} & \multicolumn{1}{|c|}{$g_{\pi^{0}\gamma\gamma}%
    \,[$GeV$]^{-1}$} & $0.273$ & $0.274\pm0.010$\\\cline{2-4}%
    \multicolumn{1}{|c}{} & \multicolumn{1}{|c|}{$\tau_{\pi^0\rightarrow\gamma\gamma}%
    \,[$s$]$} & $8.71\times10^{-17}$ & $8.57\times10^{-17}$\\\hline\hline
    \multicolumn{1}{|c}{} & \multicolumn{1}{|c|}{$|\mathcal{T}_{\eta\rightarrow\gamma
    \gamma}|\,[$eV$]^{-1}$} & $2.54\times10^{-11}$ & $(2.5\pm0.06)\times10^{-11}%
    $\\\cline{2-4}%
    \multicolumn{1}{|c}{$\eta$} & \multicolumn{1}{|c|}{$\Gamma_{\eta\rightarrow\gamma
    \gamma}\,[$KeV$]$} & $0.440$ & $0.465$\\\cline{2-4}%
    \multicolumn{1}{|c}{} & \multicolumn{1}{|c|}{$g_{\eta\gamma\gamma}%
    \,[$GeV$]^{-1}$} & $0.278$ & $0.260$\\\cline{2-4}%
    \multicolumn{1}{|c}{} & \multicolumn{1}{|c|}{$\tau_{\eta\rightarrow\gamma\gamma}%
    \,[$s$]$} & $1.52\times10^{-18}$ & $1.43\times10^{-18}$\\\hline
    \end{tabular} 
\end{center}
\end{table}

Some comments are in order concerning our calculation of the integral $I_H^i(P)$ (\ref{i3}). The fermionic action of NJL model (\ref{action}) has ultraviolet divergences and requires
a regularizing cutoff. However, the integral $I_H^i(P)$ (\ref{i3}) is not a divergent quantity. 
We chose to regularize the action from the beginning, so that there is no need to cut 
the integral $I_H^i$ -- only the ultraviolet integrals $I_{1}^i$ (\ref{I1}) and $I_2^{ij}$ (\ref{I2}) 
are regularized \cite{ruivoI}. 
The advantages of using $\Lambda\rightarrow \infty$ in non divergent integrals was already 
shown in \cite{tdavid,tklev,ruivoI}, where a better agreement with experiment of several observables 
is obtained with this procedure.

Now let us discuss our results at finite density. We  consider here the case of asymmetric quark matter imposing 
the condition of $\beta$ equilibrium  and charge neutrality through the following 
constraints, respectively on the chemical potentials and densities of quarks and electrons: $\mu_{d}=\mu_{s}=\mu_{u}+\mu_{e}\,\, \mbox{ and }\,\,\,\frac{2}{3}\rho_{u}
-\frac{1}{3}(\rho_{d}+\rho_{s})-\rho_{e}=0$, with 
$\rho_{i}=\frac{1}{\pi^{2}}(\mu_{i}^{2}-M_{i}^{2})^{3/2}\theta(%
\mu_{i}^{2}-M_{i}^{2})\,\,\mbox{ and }\,\,\,\rho_{e}=\frac{\mu_{e}^{3}}{3\pi^{2}}$ 
\cite{costaI,costaB,costabig}.

As discussed by several authors, this  version of the NJL model exhibits a first order 
phase transition \cite{RSP,costaI,buballa}. As shown in \cite{buballa}, by using a convenient 
parameterization \cite{RKH} the model may be interpreted as having a mixing phase --- droplets of light 
$u\,,d$ quarks at a critical density $\rho_c=2.25 \rho_0$ (where $\rho_0=0.17$fm$^{-3}$ is the nuclear matter density) surrounded by a non--trivial vacuum --- and, 
above this density, a quark phase with partially restored chiral symmetry \cite{costaI,buballa}.
An interesting feature of quark matter in weak equilibrium is that at densities above $\rho_s\sim 3.8\rho_0$ the mass of the strange quark becomes  lower than the chemical potential  what implies the occurrence   of strange quarks in this regime and  this fact leads to meaningful effects on the behavior of meson observables \cite{costaI,costaB,costabig}.

%%%%%%%%%%%%%%%%%%%%%%%%%%%%%%%%%%%%
\begin{figure}[t]
    \begin{center}
    \hspace{-2cm}    \includegraphics[width=0.50\textwidth]{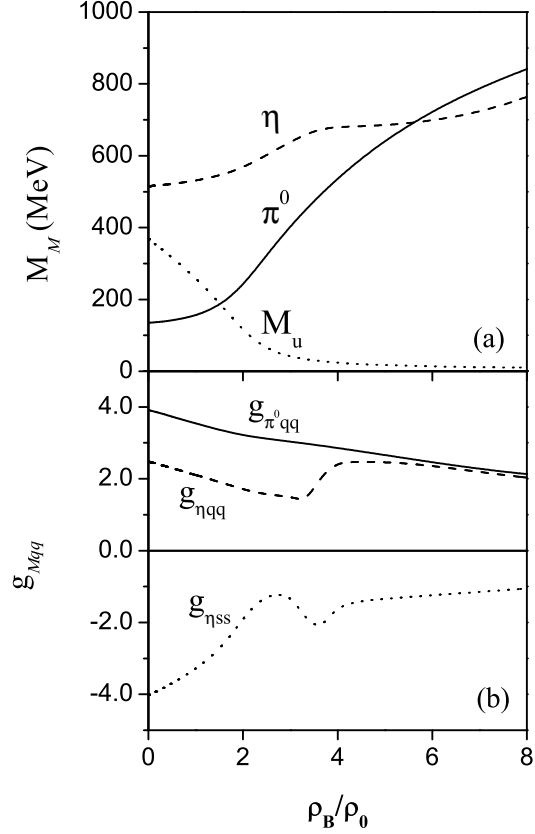}
    \end{center}
    \caption{Mesonic and quark masses (a) and meson--quark--quark coupling constant 
    				(b) as function of the baryonic density.}
    \label{fig:acopmass}
\end{figure}
%%%%%%%%%%%%%%%%%%%%%%%%%%%%%%%%%%%%

In order to evaluate the transition amplitude of the decay $H \rightarrow\gamma\gamma$ in 
function of the density, we require the behavior of $M_{H}$ and $g_{H\overline{q}q}$ 
with density, that are plotted in Fig. 2 a)-b), respectively. As we will discuss in the sequel, this behavior is essentially a manifestation of the partial restoration of chiral symmetry.  The difference between the behavior  at $T=0\,,\rho\not=0$ and $T\not=0\,,\rho=0$  is that, in the last case the mesons are no more bound states above the critical point since they dissociate in $\bar q q$ pairs at the Mott temperature.  At finite density they continue to be bound states but with a weaker coupling to the quarks as it can be seen in Fig. 2 b). So,  it is natural to expect that the effect of density on  $H\rightarrow \gamma\gamma$ decay observables be qualitatively similar to the effect of  finite temperature.  

%%%%%%%%%%%%%%%%%%%%%%%%%%%%%%%%%%%%
\begin{figure}[t]
    \begin{center}
    \hspace{-1.25cm}    \includegraphics[width=0.50\textwidth]{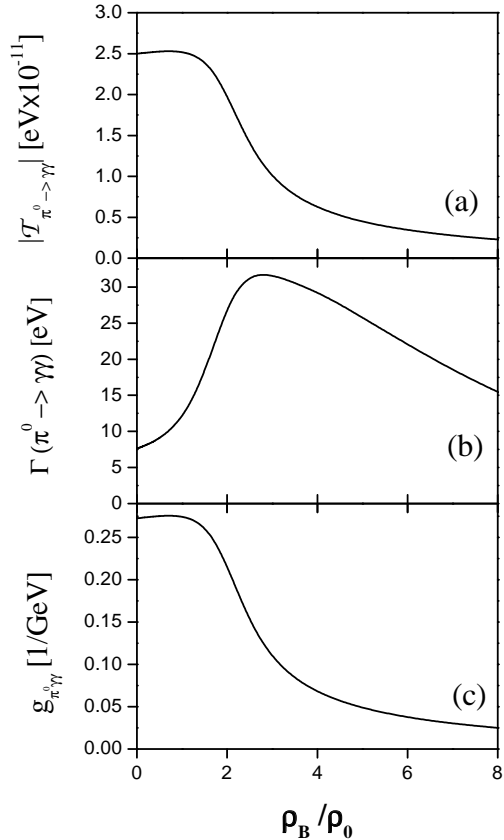}
    \end{center}
    \caption{The decay $\pi^0\rightarrow\gamma\gamma$: a) Transition amplitude; b) Decay width; c) Coupling constant.}
    \label{fig:pi0}
\end{figure}
%%%%%%%%%%%%%%%%%%%%%%%%%%%%%%%%%%%%

We discuss now our results for the medium effects on the two photon decay of $\pi^0$: 
${\mathcal T}_{\pi^0\rightarrow\gamma\gamma}\,,\Gamma_{\pi^0\rightarrow\gamma\gamma}$ and $g_{\pi^0\gamma\gamma}$ 
that are plotted in Fig. 3 a)-c). In order to understand this results let us remember the central role of this meson in connection with the breaking and restoration of chiral symmetry in the $SU(2)$ sector.  A sign for the restoration of this symmetry is that the mass of the pion increases with density and this meson becomes degenerated with $\sigma$ meson and the pion decay constant $f_\pi$ goes asymptotically to zero. 
The behavior of $\pi^0 \rightarrow \gamma\gamma$ observables with density is closely related with the restoration of chiral symmetry in the $SU(2)$ sector.  The fact that ${\mathcal T}_{\pi^0 \rightarrow \gamma\gamma}$, as well as $g_{\pi^0 \rightarrow \gamma\gamma}$  decrease with density reflects the fact that $M_u$ and $g_{\pi^0 \bar q q}$ decrease with density. As it can be seen  from Fig. 2, the quark mass decreases sharply while $g_{\pi^0\bar u u}$ decreases more slowly. In the region $\rho < \rho_c$ the behavior of the transition amplitude is dictated by a compromise between the behavior of the mass and the meson quark coupling. Above $\rho_c$  the mass decrease seems to be the dominant effect. Concerning $\Gamma_{\pi^0 \rightarrow \gamma\gamma}$  it has a maximum at about the critical density, since there are two competitive effects, on one side the decrease of the transition amplitude and, on the other side, the increase of the pion mass. Above the critical density the two photon decay of the pion becomes less  favorable, what reflects that it turns a weaker bound $\bar q q$ pair, as already mentioned before. This is compatible with recent experimental results indicating that pionic degrees of freedom are less relevant at high densities \cite{phenix}.

In order to clarify the connection between the  behavior of the anomalous couplings and the restoration of chiral symmetry one can do the simple exercise of calculating the transition amplitude in the chiral limit, by setting the external momenta equal to zero. Then, similarly to \cite{tpisarski,pisarski}, where such analysis was done with temperature, we get  $|\mathcal{T}_{\pi^{0}\rightarrow\gamma\gamma}|\propto\frac{{M_u}^2}{f_\pi}\frac{1}{{\lambda_u}^2}$, the mass decreases being the dominant effect, which leads to a vanishing of the transition amplitude at the critical density.

%%%%%%%%%%%%%%%%%%%%%%%%%%%%%%%%%%%%
\begin{figure}[t]
    \begin{center}
     \hspace{-1.25cm}   \includegraphics[width=0.50\textwidth]{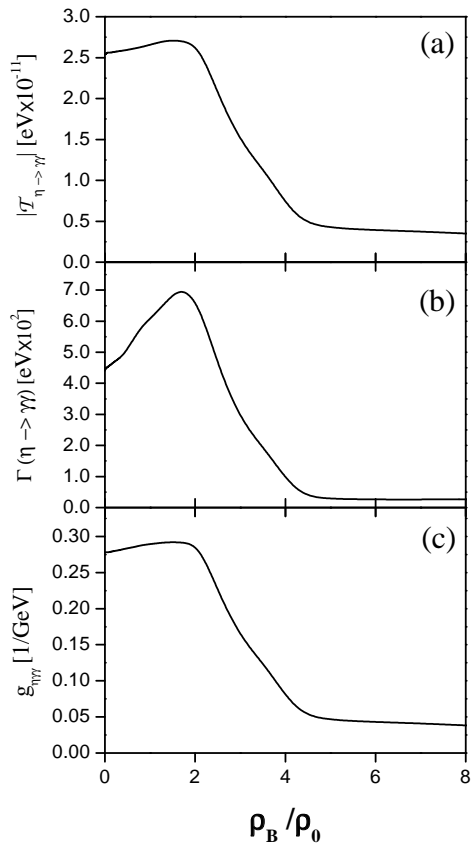}
    \end{center}
    \caption{The decay $\eta\rightarrow\gamma\gamma$: a) Transition amplitude; b) Decay width; c) Coupling constant.}
   \label{fig:eta}
\end{figure}
%%%%%%%%%%%%%%%%%%%%%%%%%%%%%%%%%%%%

Concerning the $\eta \rightarrow \gamma\gamma$ decay, although qualitatively similar to $\pi^0 \rightarrow \gamma\gamma$, there are differences that we will examine in detail and that are related to the evolution of the strange quark content of this meson and the behavior of the strange quark in this regime.  The quark content of $\eta$ is given by:
\begin{eqnarray}\label{mix} 
|\eta> &=& \cos \theta \frac{1}{\sqrt{3}} |\bar u u + \bar d d - 2 \bar s s> - \sin \theta \sqrt{\frac{2}{3}} 
|\bar u u + \bar d d + \bar s s>.
\end{eqnarray}
It was show in \cite{costaB} that  the mixing angle ($\theta=-5.8^o$ in the vacuum) decreases with density, has a minimum  $(\simeq - 25 ^0$) at $\rho\simeq 2.8 \rho_0$,  equals to zero at  $\rho\simeq 3.5\rho_0$, then it increases rapidly  up to the value $\sim 30^o$,  when strange valence quarks appear in the medium ($\rho\simeq 3.8\rho_0$). $g_{\eta \bar s s }$ (Fig. 1-b) reflects this evolution of the strange quark content. So, one sees that at high densities the  $\eta$ is governed by the behavior of the strange quark mass. Since after $\rho\simeq 3.8 \rho_0$ there is a tendency to the restoration of chiral symmetry in the strange sector, although less pronounced then for non strange quarks, this effect should manifest itself in the behavior of $\eta$ observables. Therefore, although the $\eta$ is less "Goldstone like" than the pion, one expects results qualitatively similar for the two photon decay, which are shown in Fig. 4 a)-c). The main difference is that the width $\Gamma_\eta \rightarrow\gamma\gamma$ almost vanishes above $\rho=3.8\rho_0$. This behavior of the $\eta$ seems to indicate that the role of the $U_A(1)$anomaly for the mass of this meson at high densities is less important.

In conclusion, we studied the behavior of two photon decay observables for the neutral mesons $\pi^0$ and $\eta$ in quark matter in weak equilibrium and we discussed the results in connection with restoration of symmetries.   We have shown that, in spite of the different quark structure  of these mesons,  at high densities they share a behavior which is mainly   a manifestation of restoration of chiral symmetry.  We show that these anomalous decays are significantly affected by the medium, however the relevance of this results from the experimental point of view should be discussed.  Recent experimental results from PHENIX \cite{phenix} show that $\pi^0$ production is suppressed in the central region of Au$+$Au collisions as compared to the peripherical region. This means that $\pi^0 \rightarrow \gamma\gamma$ decay could only be interesting for experimental heavy-ion collisions   at intermediate densities. However, although the peak of the $\pi^0$ width is at a moderate density (see Fig. 3) its life time, $\tau$, is here of the order of $2.08\times10^{-17}$s, much longer than the expected lifetime of the fireball in the hadronic phase, $10^{-22}$s, so the decay  should occur outside of the fireball. The same considerations apply to the $\eta$, although its maximum lifetime is $9.36\times10^{-19}$s. However, since the physics underlying these processes is the same of other anomalous processes interesting from the experimental point of view, the modification of anomalous mesonic interactions by the medium might, in principle, be observed.

%%%%%%%%%%%%%%%%%%%%%%%%%%%%%%%%%%%%%%%%%%%%%%%%%%%%%%%%%%%%%%%%%%%%%%%%%%%%%%%%%

\vspace{2cm}
\begin{center}
{\large Acknowledgment:}
\end{center}
Work supported by grant SFRH/BD/3296/2000 (P. Costa), by grant RFBR 03-01-00657, 
Centro de F\'{\i}sica Te\'orica and GTAE (Yu. Kalinovsky).

%%%%%%%%%%%%%%%%%%%%%%%%%%%%%%%%%%%%%%%%%%%%%%%%%%%%%%%%%%%%%%%%%%%%%%%%%%%%%%%%%

%%%%%%%%%%%%%%%%%%%%%%

\end{document}